\documentclass[twocolumn,aps,prl,superscriptaddress]{revtex4}
\usepackage{amsmath}

\usepackage{epsfig}
\usepackage{graphicx}
\usepackage{bm}
\usepackage{amssymb}%

\def\gapp{\lower.35em\hbox{$\stackrel{\textstyle>}{\sim}$}}
\def\lapp{\lower.35em\hbox{$\stackrel{\textstyle<}{\sim}$}}

\begin{document}

\bibliographystyle{apsrev}

\title{Tunable Casimir repulsion with three dimensional topological insulators}

\author{Adolfo G. Grushin}
\affiliation{Instituto de Ciencia de Materiales de Madrid (CSIC),\\
Sor Juana In\'{e}s de la Cruz 3, Madrid 28049, Spain.}
\author{Alberto Cortijo}
\affiliation{Department of Physics, Lancaster University, Lancaster, LA1 4YB, United Kingdom}

\begin{abstract}
In this Letter, we show that switching between repulsive and attractive Casimir forces by means of external tunable parameters could be realized with two topological insulator plates. We find two regimes where a repulsive (attractive) force is found at small (large) distances between the plates, canceling out at a critical distance. For a frequency range where the effective electromagnetic action is valid, this distance appears at length scales corresponding to $1-\epsilon(\omega)\sim \frac{2}{\pi}\alpha\theta$.
\end{abstract}

\maketitle
The full experimental accessibility to micrometer and sub-micrometer size physics and the possibility of developing applications has turned the understanding of phenomena at these scales to be of fundamental importance. Within this scenario, the Casimir force \cite{Cas48} arises when two objects are placed near each other at distances of a few micrometers. In the general case of two dielectrics the situation is well described by the theory developed by Dzyaloshinskii \emph{et. al} \cite{DLP61} where the optical response of the material determines the magnitude and behaviour of the force. In the simplest case of a mirror symmetric situation a theorem ensures that the force is always attractive \cite{K06,B06}, resulting in a problem for nano-mechanical devices. To revert the sign, one must search non symmetric situations, usually adding complexity to the problem.
The first Casimir repulsion proposal, known as Dzyaloshinskii repulsion, was recently confirmed experimentally  \cite{MCP09} and it involves a third dielectric medium between the plates, excluding the possibility of frictionless devices and quantum levitation. In turn, vacuum mediated proposals include magnetic versus non-magnetic situations \cite{B74} and the use of metamaterials \cite{RDM06,LP07,ZZK09a}.
\begin{figure}
\begin{center}
\includegraphics[scale=0.12]{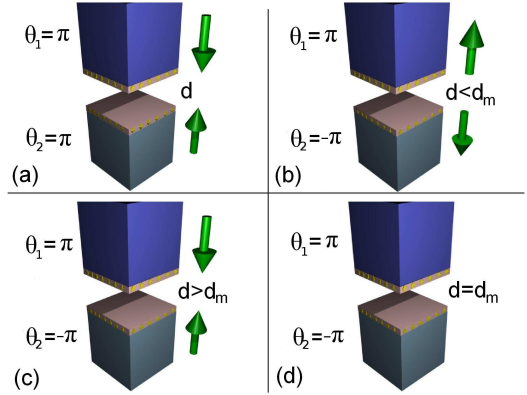}
\end{center}
\caption{\label{Fig:exp} Different configurations of the Casimir effect with identical TI covered with a thin magnetic layer.(a) The magnetization is of the same sign on each surface resulting in a case where $\theta_{1}=\theta_{2}$ giving Casimir attraction. In (b), (c) and (d) the magnetizations have opposite signs on the surface ($\theta_{1}=-\theta_{2}$) leading to attraction when $d>d_{m}$, repulsion when $d<d_{m}$ and to a quantum levitation configuration at $d_{m}$ where the net force is zero.}
\end{figure}
In this Letter, we report a new method for obtaining a twofold tunable Casimir repulsion. By use of the optical properties of topological insulators (TI) it is feasible to achieve all situations between repulsion to attraction by using two controllable parameters: the distance between dielectrics, and the sign of the topological magnetoelectric polarizability (TMEP) $\theta$, where the latter allows to tune the optical properties of the mentioned materials.\\
TIs are characterized by a bulk insulating behaviour with metallic boundary states protected by time reversal symmetry \cite{KM05b,QHZ08}. 
The topological protection of edge states ensures their stability against non-magnetic perturbations. The $3$D counterpart of this novel topological state was shown to exist in a Bi$_{x}$Sb$_{1-x}$ alloy \cite{H08} and in the stoichiometric crystals Bi$_{2}$Se$_{3}$, Bi$_{2}$Te$_{3}$,TlBiSe$_{2}$ and Sb$_{2}$Te$_{3}$ \cite{C09,X09,Chen10}. \\
The Casimir force is intimately related to the optical properties of the two dielectric bodies \cite{DLP61}. For instance, consider the situation where two dielectric parallel semi-infinite bodies (labeled $1$ and $2$) are placed at a distance $d$ from each other in vacuum. In this case, the Casimir energy density (CED) stored by the plates is given by \cite{Reyn91}:
\begin{equation}
\label{CasimirEnergy}
\frac{E_{c}(d)}{A\hbar} = \int_0^{\infty} \hspace{-1pt} \frac{d\xi}{2\pi} \int \frac{d^2 {\bf k}_{\|}}{(2\pi)^2} \log \det \left[1 - {\bf R}_1 \cdot {\bf R}_2 e^{-2 k_3 d}\right] ,
\end{equation}
where $A$ is the plate area, $k_3=\sqrt{\bm{k}^{2}_{\|}+ \xi^2/c^2}$ is the wave vector perpendicular to the plates, $\bm{k}_{\|}$ is the vector parallel to the plates and $\xi$ is the imaginary frequency defined as $\omega = i\xi$. The matrices ${\bf R}_{1,2}$ are $2\times2$ reflection matrices of media $1$ and $2$ containning the Fresnel coefficients defined as:
\begin{eqnarray}
\label{ReflectionMatrices}
{\bf R} = \left[
\begin{array}{cc}
   R_{s,s} (i \xi, {\bf k}_{\|}) &  R_{s,p} (i \xi, {\bf k}_{\|}) \\
  R_{p,s} (i\xi, {\bf k}_{\|}) &  R_{p,p} (i \xi, {\bf k}_{\|})
\end{array} \right] ,
\end{eqnarray}
where $R_{i,j}$ describes the reflection amplitude of an incident wave with polarization $i$ which is reflected with polarization $j$. The polarizations $R_{s}$ ($p$) describe parallel (perpendicular) polarization with respect to the plane of incidence. The Casimir force per unit area on the plates is obtained by differentiating expression (\ref{CasimirEnergy}). A positive (negative) force, or equivalently a positive (negative) slope of $E_{c}(d)$, corresponds to attraction (repulsion) of the plates. \\
The electromagnetic response of a dielectric, which defines the reflection matrices, is governed by Maxwell's equations derived from the ordinary electromagnetic action $S_{0} = \int dx^3 dt \left(\epsilon \bm{E}^2-(1/\mu)\bm{B}^2\right)$, being $\bm{E}$ and $\bm{B}$ the electric and magnetic fields respectively. TI in three dimensions are well described by adding a term of the form $S_{\theta} = (\alpha /4\pi^2 )\int dx^3 dt \hspace{1mm} \theta\hspace{1mm} \bm{E}\cdot\bm{B}$, where $\alpha=1/137$ is the fine structure constant and $\theta$ is the TMEP (axion field) \cite{QHZ08,EMB09}. Because of time reversal symmetry, this term is a good description of the bulk of a trivial insulator (e.g. vacuum) when $\theta=0$ and of the bulk of a TI when $\theta=\pi$. However, the axion coupling is only a good description of both the bulk and the boundary of a TI when a time reversal breaking perturbation is induced on the surface and the system becomes fully gapped. In this situation, $\theta$ can be shown to be quantized in odd integer values of $\pi$ such that $\theta =(2n + 1)\pi$ where $n \in \mathbb{Z}$. The value of $n$ is determined by the nature of the time reversal breaking perturbation, which could be controllable experimentally by covering the TI with a thin magnetic layer. In particular, positive or negative values of $\theta$ are related to different signs of the magnetization on the surface \cite{QLZZ09}. As we will demonstrate in what follows, the Casimir force is very sensitive to the value of $\theta$ and the tunability of its sign will allow us to describe a mechanism for switching between repulsive and attractive forces,\\
\begin{figure*}
\begin{minipage}{.49\linewidth}
\includegraphics[scale=0.45]{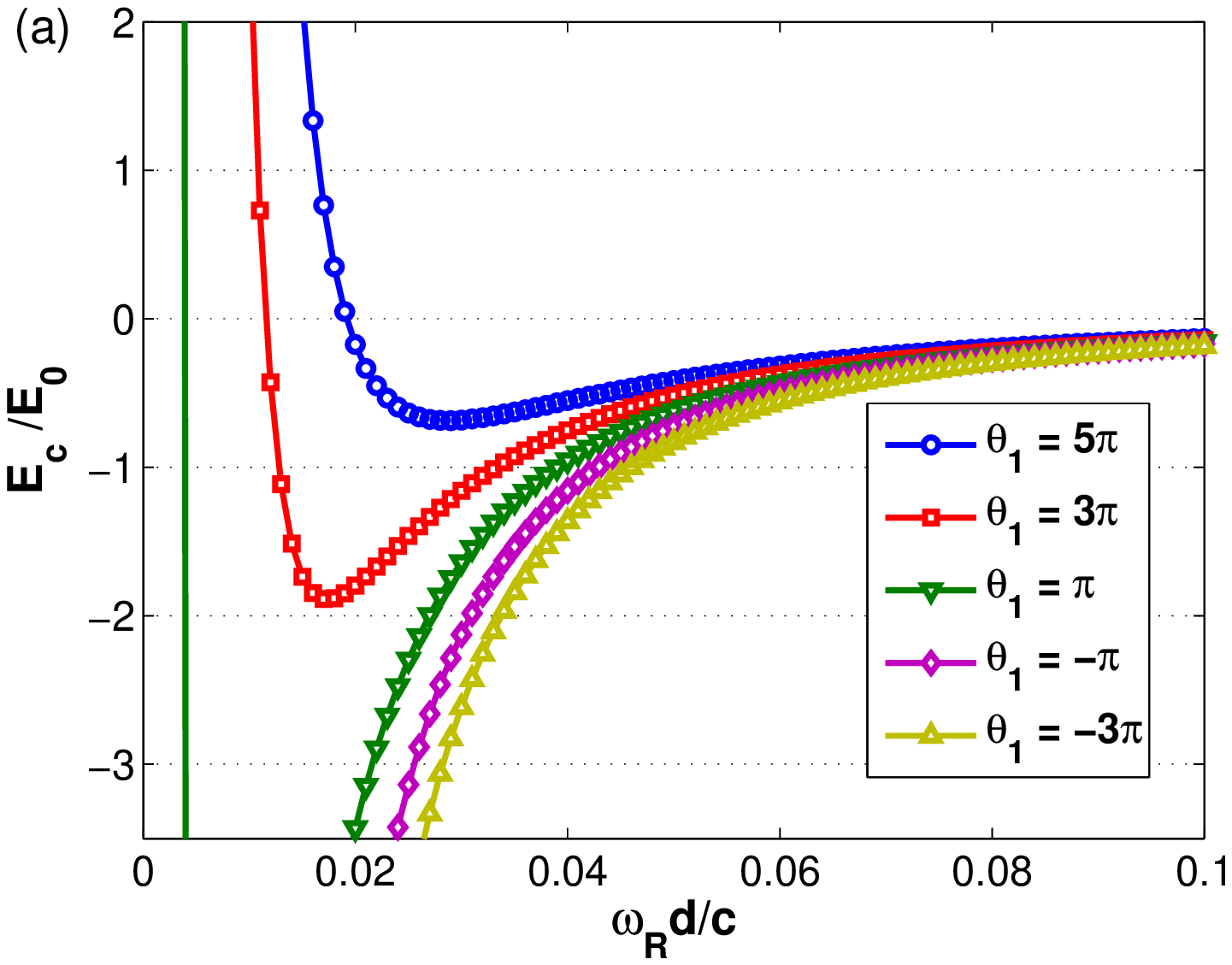}
\end{minipage}
\begin{minipage}{.49\linewidth}
\includegraphics[scale=0.45]{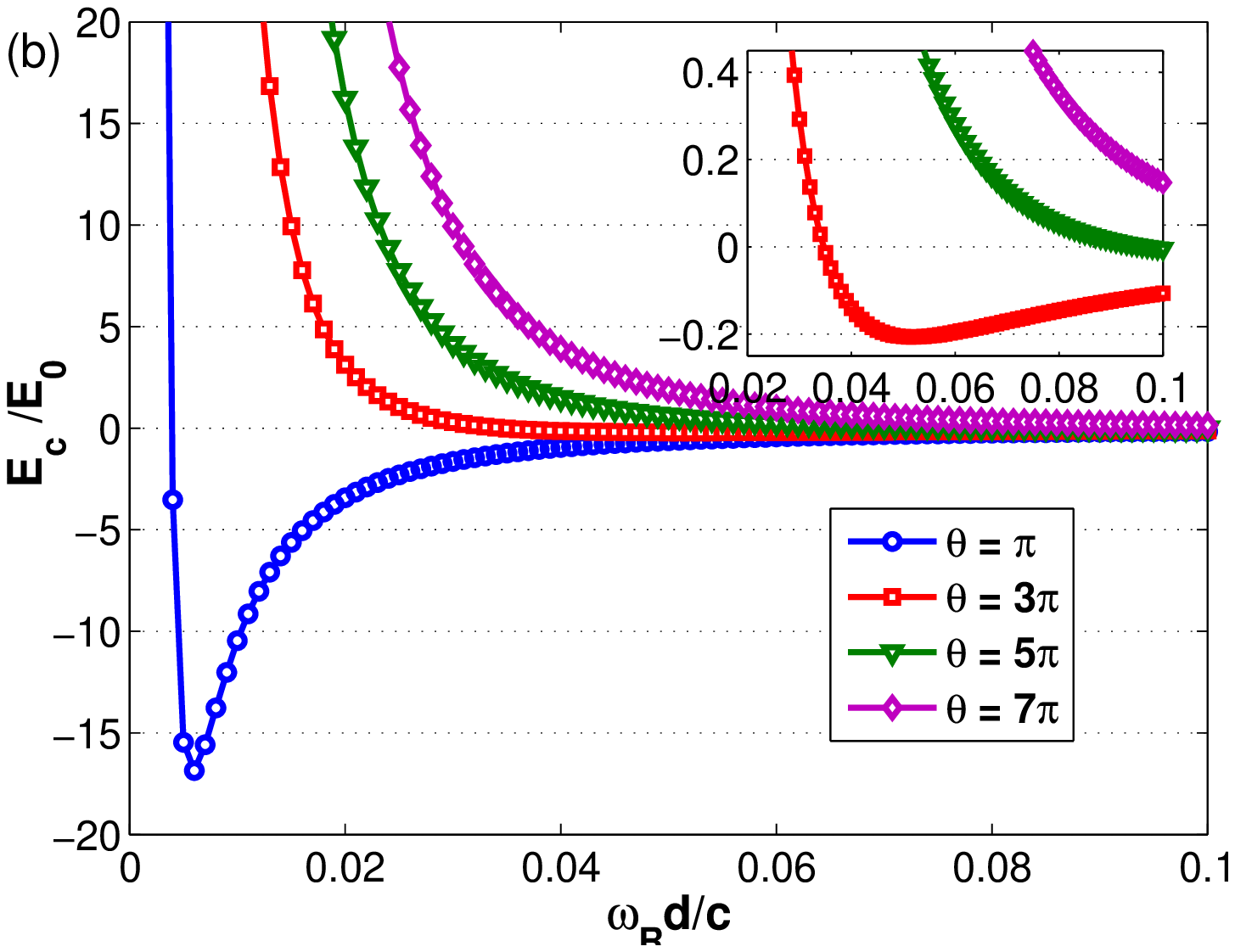}
\end{minipage}
\caption{\label{Fig:alphaeff}Casimir energy density (in units of $E_{0}= A\hbar c/(2\pi)^2(\omega_{R}/c)^3$) as a function of the dimensionless distance $\bar{d}$ for $\omega_{e}/\omega_{R}=0.45$. In (a) $\theta_{2}=-\pi$ is fixed. Whenever $\mathrm{sgn}(\theta_{1}) = -\mathrm{sgn}(\theta_{2})$ a minimum $\bar{d}_{m}$ appears leading to a vanishing net force on the plates. Increasing $\theta_{1}$ within positive values suppresses the minimum shifting $\bar{d}_{m}$ towards lower values (if both signs are equal then only attractive behaviour occurs). Complete repulsion is achieved when one of the TMEP is much bigger than the other. (b) The optimal situation $\theta_{1}=-\theta_{2}=\theta$. Different values of $\theta$ show that the minimum is enhanced when the difference between the two values is as small as possible ($\theta=\pi$).}
\end{figure*}
The electromagnetic response of a system in the presence of a $\theta$ term is still described by the ordinary Maxwell equations but the constituent relations which define the electric displacement $\bm{D}$ and the magnetic field $\bm{H}$ acquire an extra term proportional to $\theta$ \cite{Wilczek87} $\bm{D} = \epsilon\bm{E}+\alpha(\theta/\pi)\bm{B}$ and $\bm{H} = \bm{B}/\mu-\alpha(\theta/\pi)\bm{E}$. We note that eq.(1) can be easily modified to take into account magnetoelectric couplings (this happens also in chiral metamaterials\cite{ZZK09a}). The result is the same equation with the proper reflection matrices.
It is then possible to derive by means of ordinary electromagnetic theory the reflection coefficients of a TI-vacuum interface. For a TI characterized by a frequency dependent dielectric function $\epsilon(\omega)$ and a TMEP $\theta$ the reflection coefficients will take a symmetric form \cite{CY09} where the off diagonal coefficient can be expressed as:
\begin{equation}
\label{eq:ReflectionMatricesTI}
R_{s,p} (i \xi, {\bf k}_{\|},\theta) = \mathrm{sgn}(\theta) r_{sp}(i \xi, {\bf k}_{\|},\theta),
\end{equation}
where $r_{sp}$ is an even function of $\theta$. When $\theta=0$, $R_{s,p}=0$, leading to the usual attractive Casimir force due to the non-mixing of polarizations \cite{KMM09}. When $\theta \neq 0$ the reflection coefficients mix polarizations and  the sign of $\theta$ plays a crucial role on the sign of the Casimir force (see below). \\
In what follows we will consider that the surface of the TIs of the Casimir system are covered by a thin magnetic layer as shown in Fig.1. This effectively turns the TI into a full insulator (both in the bulk and on the surface) which can be safely described with the TMEP and a dielectric function, as shown in earlier works \cite{QHZ08,QLZZ09}. Hence, to numerically compute the CED by means of (\ref{CasimirEnergy}) a model for the dielectric function is necessary (we henceforth assume $\mu=1$). Due to the low concentration of free carriers in insulators the most general phenomenological model to describe the optical response of a dielectric is a sum of oscillators to account for particular absorption resonances \cite{KMM09}. When only one oscillator is considered (see however \cite{AI}) the dielectric function evaluated can be written as:
\begin{equation}\label{eq:phendiel}
\epsilon(i\xi)=1+\dfrac{\omega_{e}^2}{\xi^2+\omega^2_{R}+\gamma_{R}\xi} ,
\end{equation}
In this model, $\omega_{R}$ is the resonant frequency of the oscillator while $\omega_{e}$ accounts for the oscillator strength. The damping parameter $\gamma_{R}$ satisfies $\gamma_{R}<<\omega_{R}$ playing therefore a secondary role. In what follows, we have rescaled all quantities in units of $\omega_{R}$ leaving the quantity $\epsilon(0) \equiv 1 + (\omega_{e}/\omega_{R})^2 $  as the only parameter of the model. A good candidate to be described by this model is the TI TlBiSe$_{2}$\cite{Chen10}. This material has experimentally \cite{Mit92} (neglecting free carrier contributions and assuming high frequency transparency) $\epsilon(0)\sim 4$ and has a single resonant frequency near $56$cm$^{-1}$. Other TI could need more oscillators to be added in (\ref{eq:phendiel}).\\
We have computed the CED between two TI plates described by the TMEP $\theta_{1}$ and $\theta_{2}$ and the value of the dielectric constant at zero frequency $\epsilon(0)$ by numerical evaluation of expression (\ref{CasimirEnergy}). The results are summarized in figures \ref{Fig:alphaeff} and \ref{Fig:Aeff} where the CED is plotted against the dimensionless distance $\bar{d} \equiv d \omega_{R}/c$. From Fig. \ref{Fig:alphaeff} (a) it is clear that opposite signs of $\theta_{1,2}$ lead to the existence of a minimum ($\bar{d}_{m}$) where the net force is zero. The behaviour is attractive when both signs become equal suggesting that it is possible to tune the Casimir force by tuning the relative signs of $\theta$, i.e switching the magnetizations of the coverings. The existence of a minimum is analytically shown below in terms of the relative importance of the off-diagonal terms (\ref{eq:ReflectionMatricesTI}) against the diagonal terms and in terms of the relative sign of the TMEPs.  At large distances the diagonal terms dominate and the usual Casimir attraction is recovered. At small distances, the off-diagonal terms dominate and their sign determines whether the CED approaches $\pm\infty$ (i.e. repulsive or attractive) leading to a minimum at intermediate distances if the signs of $\theta_{1,2}$ are opposite. \\
To prove the existence of the minimum we consider the Fresnel equations for TI obtained earlier in \cite{CY09} which lead to eq. (\ref{eq:ReflectionMatricesTI}) added to the following properties of the dielectric function: finite dielectric permittivity at zero frequency ($\epsilon(0)<\infty$) and high frequency transparency,  $\epsilon(\omega)\rightarrow 1 $ when $\omega\rightarrow \infty$. For analytical traceability we assume $\theta_{1}=-\theta_{2}\equiv\theta$ and that the dielectric function (\ref{eq:phendiel}) describes the TI, although the derivation does not depend on the explicit form of the dielectric function as long as it fulfils the mentioned conditions.\\
In (\ref{CasimirEnergy}) we can rescale $\xi$ and ${\bf k}_{\|}$ to contain $\bar{d}$ which gives an overall factor $1/\bar{d}^3$ and forces the reflection matrices to be evaluated at the rescaled frequency and momenta $\xi /\bar{d}$ and ${\bf k}_{\|}/\bar{d}$. Hence, $E_{c}(\bar{d}\rightarrow 0) \rightarrow \pm \infty$ and $E_{c}(\bar{d}\rightarrow \infty) \rightarrow 0$ since the behaviour of $\epsilon(i\xi)$ ensures that the reflection matrices are not singular when evaluated at $i\xi/\bar{d} \rightarrow 0$ and $i\xi/\bar{d} \rightarrow \infty$.\\
The way the integral approaches these limits determines the sign of $E_{c}(\bar{d})$. For instance, if the integrand is positive at small distances and negative at large distances, necessarily a minimum exists at an intermediate distance $\bar{d}_{m}$. In what follows it will be shown that this is exactly what happens unless $\epsilon(0)=1$, where both limits are positive and hence long range repulsion is obtained.
Under these conditions the diagonal terms in (\ref{ReflectionMatrices}) are equal for both TI (which we label $r_{s}$ and $r_{p}$) and the off-diagonal terms given by (\ref{eq:ReflectionMatricesTI}) have opposite overall signs, but equal absolute value given by the function $r_{sp}$. Introducing these inside (\ref{CasimirEnergy}) the integrand reads:
\begin{eqnarray}\label{eq:integrand}
I = & \log[1+e^{-2 k^{(r)}_3}(2 r_{sp}^2-r_{p}^2-r_{s}^2)+\\
& + e^{-4 k^{(r)}_3}(r_{sp}^2-r_{p}r_{s})^2], \nonumber
\end{eqnarray}
where $k^{(r)}_3$ is now evaluated at the rescaled frequency and momenta just as the reflection matrices. In the limit of small distances ($\bar{d} \rightarrow 0$) and using the high frequency transparency of the dielectric function it can be shown that $|r_{s}|,|r_{p}|<<|r_{sp}|$ since the first are of order $\alpha^2$ and the second are of order $\alpha$.  Hence the integrand is positive and so $E_{c}(\bar{d}\rightarrow 0) \rightarrow +\infty$. \\
Now we consider the limit of large distances ($\bar{d} \rightarrow \infty$). In this limit, the reflection coefficients take the form $r_{s} = \left(1-\epsilon (0)-\bar{\alpha}^2\right)/D$ (a similar expression holds for $r_{p}$) and $r_{sp} =  2|\bar{\alpha}|/D$ where $D = 1+\epsilon(0)+\bar{\alpha}^2 + \sqrt{\epsilon(0)}\chi $ , $\chi$ is a frequency and momentum dependent function irrelevant for the present discussion and $\bar{\alpha}=\alpha\theta/\pi$. In this long distance limit, depending on the values of $\epsilon(0)$ different behaviors emerge. Since $\epsilon(0) \geq 1$ we now consider the two extreme limits, one where $\epsilon(0)=1$ and the other with $\epsilon(0)>>1$. \\
In the limit where $ \epsilon(0) >> 1$, the condition $|r_{s}|,|r_{p}|>>|r_{sp}|$ is always satisfied. When $\epsilon(0)$ is strictly infinity we recover the ideal case of an ordinary metal with $r_{s,p} = \pm 1$ and $r_{sp} = 0$. The integrand at large distances is a negative quantity and so $E_{c}(\bar{d})$ approaches zero from negative values. From the previous discussion at small distances $E_{c}(\bar{d})\rightarrow +\infty$, therefore there must be a minimum at an intermediate distance $ 0 < \bar{d}_{m} < \infty$ since the function must cross the $x$ axis. In the unrealistic case where $\epsilon(0)=1$ one can check that $|r_{sp}|>>|r_{s}|,|r_{p}|$ making $E_{c}(\bar{d})$ always positive for all distances. In this case there is no minimum and the force is always repulsive.
\begin{figure}
\includegraphics[scale=0.5]{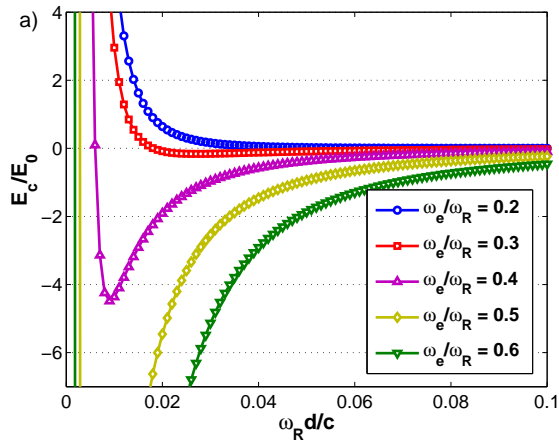}
\caption{\label{Fig:Aeff} Effect of parameter $\epsilon(0)$ with fixed $\theta_{1}=-\theta_{2}= \pi$. The effect of increasing $\epsilon(0)$ is to develop a minimum, which shifts to smaller $\bar{d}_{m}$ as $\epsilon(0)$ is increased.}
\end{figure}
By this analytical analysis and when $\theta_{1}=-\theta_{2}$ we expect that as we increase $\epsilon(0)$ from one, a minimum develops at an intermediate distance $\bar{d}_{m}$. This distance shifts to lower values as we increase $\epsilon(0)$ until, at $\epsilon(0) = \infty$ we recover the metallic case where complete attraction occurs. In the case where $\theta_{1}=\theta_{2}$ the signs inside (\ref{eq:integrand}) change and make the logarithm to be negative recovering attraction at all intermediate distances (for more details see \cite{AI}). The consistency of these analytical expectations is confirmed numerically with the results shown in figures \ref{Fig:alphaeff} and \ref{Fig:Aeff}.\\
From these we infer that in order to enhance as much as possible the minimum, it is necessary to search for a situation where $\theta \equiv \theta_{1}=-\theta_{2}$. The CED for different values of $\theta$ satisfying this condition are depicted in figure Fig. \ref{Fig:alphaeff} (b). Under these circumstances the minimum is more prominent when $\theta= \pi$, i.e. when $\theta$ takes its smallest possible value. The general analytical analysis, supported by the numerical evaluation for different parameters, suggest that a simple experimental set-up (Fig. \ref{Fig:exp}) could switch from complete Casimir attraction to a stable quantum levitation regime by reversing the magnetization of one of the layers covering one of the TI. In this process the system will turn from a symmetric situation where $\theta_{1}=\theta_{2}$  resulting in attraction to a non-symmetric situation where the optimum condition is satisfied ($\theta_{1}=-\theta_{2}$) and a stable minimum appears.\\
To conclude, for this appealing situation to be experimentally accessible one has to search for realizations of $\omega_{R}$ where the typical distances between the plates are at least of order $0.1\mu$m - $1\mu$m. For a frequency range where the axion lagrangian is valid \cite{EMB09}, the minimum is expected to appear at a position where diagonal and off-diagonal terms are similar in magnitude, i.e. length scales corresponding to $1-\epsilon(\omega)\sim \frac{2}{\pi}\alpha\theta$. Low values of $\epsilon(0)$ (typically less than 10) favor this situation since the minimum is realized at larger distances. While the electromagnetic parameters of TI are still not well characterized, a low $\epsilon(0)$ could be achieved by using thin films or by air injection which will lower the bulk dielectric response. For TlBiSe$_{2}$ we estimate from numerical integration that the minimum of the CED appears at a distance of $d=0.1\mu$m and with a CED of the same order than for the usual metal-vacuum-metal system at $1\mu$m, hence being still experimentally accessible. We must note however that this estimation requires a high TMEP value ($\theta \sim 10\pi$) in order to shift the minimum to observable distances. Therefore the proposed effect is on the verge of experimental accessibility and should encourage experimental efforts to attain full optical characterization of TI. We stress here that the only effect of the magnetic coating is to gap the surface states. We have estimated the parasitic magnetic forces between the magnetic layers following \cite{Bruno02}. The dipole-dipole interaction is of the order of attoN at distances of 50 nm and the magnetic Casimir force\cite{Bruno02} is $\sim 1 fN$, much smaller than the force described here which is of the order of 5pN. The proposed effect could also be explored in other magnetodielectric materials such as Cr$_{2}$O$_{3}$ which can be described by a higher axion coupling\cite{Obu08}. However, these materials induce more general magnetoelectric couplings \cite{EJV10} which we will consider in a future work. \\
We acknowledge F. de Juan, M.A.H. Vozmediano, F. Guinea, J. Sabio, and B. Valenzuela for very useful discussions and suggestions. A.G.G acknowledges MICINN (FIS2008-00124) and A.C. acknowledges EPRSC Science and Innovation Award(EP/G035954) for funding.

\begin{widetext}

\section*{Auxiliary Material:\\ Tunable Casimir repulsion with three dimensional topological insulators}

\maketitle

\subsection*{Attraction versus repulsion from Fresnel coefficients:}

\textit{1.Fresnel coefficients for topological insulators:}
In order to compute the Casimir energy $E_{c}(d)$, the Fresnel coefficients which define the reflection matrices of a topological insulator have to be calculated. These coefficients relate the amplitude of the incident and reflected electric fields for different polarizations. Since the normal components of $\bm{D}$ and $\bm{B}$ and tangential components of $\bm{E}$ and $\bm{H}$ must be continuous along the interface one can obtain the following reflection amplitudes by defining $\bm{H}$ and $\bm{D}$ as in the main text \cite{CY09,Obu09}:
\begin{eqnarray}
\label{eq:ReflectionMatricesTIcomplete}
\left(
\begin{array}{c}
  E^{(r)}_{s}  \\
 E^{(r)}_{p}
\end{array} \right) = \dfrac{1}{\Delta}\left(
\begin{array}{cc}
  n^2_{1}-n^2_{2}-\bar{\alpha}^2 + n_{1}n_{2}\chi_{-} &  2 \mathrm{sgn}(\theta_{2}-\theta_{1})|\bar{\alpha}| n_{1} \\
  2 \mathrm{sgn}(\theta_{2}-\theta_{1})|\bar{\alpha}| n_{1} & -n^2_{1}+n^2_{2}+\bar{\alpha}^2 + n_{1}n_{2}\chi_{-}
\end{array} \right)\left(
\begin{array}{c}
  E^{(i)}_{s}  \\
 E^{(i)}_{p}
\end{array} \right),
\end{eqnarray}
where $\bar{\alpha}=\alpha(\theta_{2}-\theta_{1})/\pi$, $\alpha$ is the fine structure constant ($\alpha = \dfrac{e^2}{\hbar c}$), $n_i$ is the refractive index of material $i$, $\Delta = n^2_{1}+n^2_{2}+\bar{\alpha}^2 + n_{1}n_{2}\chi_{+}$ and:

\begin{eqnarray}
\label{eq:Xipm}
 \chi_{\pm}= 
 \dfrac{\xi^2+\frac{\mathbf{k}_{\|}^2}{n_{1}^2} \pm \left(\xi^2 + \frac{\mathbf{k}_{\|}^2}{n_{2}^2}\right)}{\sqrt{\left(\xi^2 + \frac{\mathbf{k}_{\|}^2}{n_{1}^2}\right)\left(\xi^2 + \frac{\mathbf{k}_{\|}^2}{n_{2}^2}\right)}}.
\end{eqnarray}

When $\bar{\alpha}=0$ and after a little algebra, these coefficients reduce to the ordinary Fresnel coefficients for a dielectric-dielectric interface \cite{Born93}. Since we are considering a topological insulator-vacuum interface, $n_{1}=1$, $\theta_{1}=0$, $\theta_{2}=\theta$ and $n^2_{2}(\omega)=\epsilon(\omega)$, therefore $\bar{\alpha} = \alpha\theta/\pi$. By its definition, $\Delta$ is always positive and the off diagonal terms have a an overall sign governed by the sign of $\theta$. \\
To calculate the Casimir energy stored between the plates one computes:

\begin{equation}
\label{CasimirEnergys}
\frac{E_{c}(d)}{A\hbar} = \int_0^{\infty} \hspace{-1pt} \frac{d\xi}{2\pi} \int \frac{d^2 {\bf k}_{\|}}{(2\pi)^2} \log \det \left[1 - {\bf R}_1 \cdot {\bf R}_2 e^{-2 k_3 d}\right] ,
\end{equation}

equation (1) in the main paper, where ${\bf R}_i$ is the $2$x$2$ matrix defined in \eqref{eq:ReflectionMatricesTIcomplete}.\\

\textit{2. Analytical existence of a minimum and the role of different parameters:}
To prove the existence of the minimum it is enough to assume that the reflection coefficients are given by the expression \eqref{eq:ReflectionMatricesTIcomplete} added to the analytical properties of the dielectric function, i.e. finite dielectric permittivity at zero frequency ($\epsilon(0)<\infty$) and high frequency transparency: $\epsilon(\omega)\rightarrow 1 $ when $\omega\rightarrow \infty$. For an insulator one can assume the dielectric function:

\begin{equation}\label{eq:phendiels}
\epsilon(i\xi)=1+\dfrac{\omega_{e}^2}{\xi^2+\omega^2_{R}+\gamma_{R}\xi} ,
\end{equation}

although the derivation does not depend on the analytical form of the dielectric function as long as it fulfils the mentioned conditions. In particular, it holds for more than one oscillator. In addition, and for analytical traceability we might further assume the particular situation where $\theta_{1}=-\theta_{2}$ (where labels $1$ and $2$ identify the Casimir plates).\\

It is straightforward to check that if in the expression for $E_{c}(d)$ we rescale $\xi$ and ${\bf k}_{\|}$ to contain $d$ this gives an overall factor $1/d^3$ and forces the reflection matrices to be evaluated at the rescaled frequency and momenta $\xi /d$ and ${\bf k}_{\|}/d$. Thus all of the $d$ dependence can be transferred to the reflection matrices and the overall $1/d^3$ prefactor. The expression for the energy turns to be:

\begin{equation}
\label{CasimirEnergyres}
\frac{E_{c}(d)}{E_{0}} = \dfrac{1}{d^3}\int_0^{\infty} \hspace{-1pt} d\xi \int d^2 {\bf k}_{\parallel} \log \det \left[1 - {\mathbf{R}_1\left[\dfrac{\xi}{d},\dfrac{\mathbf{k}_{\parallel}}{d}\right]} \cdot {\mathbf{R}_2\left[\dfrac{\xi}{d},\dfrac{\mathbf{k}_{\parallel}}{d}\right]} e^{-2 k_3}\right] ,
\end{equation}

where $k^{2(r)}_{3} = \xi^2+k_{\parallel}^{2}$ is defined through the rescaled variables (which include $c$ the speed of light in vacuum and $\omega_R$ defined in the main text) and $E_{0}= A\hbar c/(2\pi)^2(w_{R}/c)^3$ where $A$ is the area of the plates and the reflection matrices are evaluated at these rescaled variables with a rescaled dielectric function:

\begin{equation}\label{eq:phendielres}
\epsilon(i\xi/d)=1+\dfrac{\left( \dfrac{\omega_{e}}{\omega_{R}}\right) ^2}{\left( \dfrac{\xi}{d}\right) ^2+ 1+\dfrac{\gamma_{R}}{\omega_{R}}\dfrac{\xi}{d}}.
\end{equation}

The behaviour of $\epsilon(i\xi)$ ensures that the reflection matrices are not singular when evaluated at $\xi/d \rightarrow 0$ and $\xi/d \rightarrow \infty$. Hence we can conclude that $E_{c}(d\rightarrow 0) \rightarrow \pm\infty$ and $E_{c}(d\rightarrow \infty) \rightarrow 0$. \\
The way the integrand approaches these limits determines the sign of $E_{c}(d)$. For instance, if the integrand is positive at small distances and negative at large distances, necessarily a minimum exists at an intermediate distance $d_{c}$. In what follows it will be shown that this is exactly what happens when $\theta_{1}=-\theta_{2}$ (unless $\epsilon(0)=1$, where both limits are positive and hence long range repulsion is obtainned).
The first step is to evaluate the integrand in \eqref{CasimirEnergys}. For the situation in which $\theta_{1}=-\theta_{2}= \theta$ the reflection matrices describing both topological insulators are:

\begin{eqnarray}
\label{ReflectionMatricesTIpm}
{\bf R}_{\pm} = \left[
\begin{array}{cc}
   r_{s} (i \xi, {\bf k}_{\|}) &  \pm r_{sp} (i \xi, {\bf k}_{\|}) \\
  \pm r_{sp} (i\xi, {\bf k}_{\|}) &  r_{p} (i \xi, {\bf k}_{\|})
\end{array} \right] .
\end{eqnarray}

Introducing this inside \eqref{CasimirEnergys} we obtain the following integrand:

\begin{eqnarray}\nonumber
I = \log \det \left[1 - \mathbf{R}_{+}\cdot\mathbf{R}_{-} e^{-2 k^{(r)}_3}\right] = \log\left[ 1 + e^{-2 k^{(r)}_3} \left(2 r_{sp}^2-r_{p}^2-r_{s}^2\right) + e^{-4 k^{(r)}_3} \left(r_{sp}^2-r_{p}r_{s}\right)^2\right],
\end{eqnarray}

Notice that the last term although always positive, will play no role in what follows since it is always suppressed over the first term (note that $r_{s},r_{p},r_{sp} < 1$).\\

In the limit of \textbf{small distances} ($d \rightarrow 0$) and using the high frequency transparency of the dielectric function it easy to show that $|r_{i}|<<|r_{sp}|$ ($i=s,p$):
Notice first that the denominator $\Delta$ defined in \eqref{eq:ReflectionMatricesTIcomplete} is common to all terms so it cannot play a role on the relative magnitude of the coefficients. We need however to study the behaviour of $\chi_{-}$ at small distances, given by:

\begin{eqnarray}
\label{eq:Ximsmall}
 \chi_{-}\left(\dfrac{\xi}{d},\dfrac{\mathbf{k}_{\parallel}}{d}\right) = 
 \dfrac{\xi^2 + \mathbf{k}_{\|}^2 - \left(\xi^2 + \frac{\mathbf{k}_{\|}^2}{n_{2}^2}\right)}{\sqrt{\left( \xi^2 + \mathbf{k}_{\|}^2\right) \left(\xi^2 + \frac{\mathbf{k}_{\|}^2}{n_{2}^2}\right)}}.
\end{eqnarray}

Remembering that at small distances (large frequencies) we have transparency, $n_{2}^{2}(\xi/d)= \epsilon(\xi/d) \rightarrow 1$ then it is straightforward to see that $\chi_{-}\rightarrow 0 $ and hence:

\begin{equation}\nonumber
r_{s} = \dfrac{-\bar{\alpha}^2}{2+\bar{\alpha}^2 + \chi_{+}} = -r_{p},
\end{equation}

and

\begin{equation}\nonumber
r_{sp} = \dfrac{2\bar{\alpha}}{2+\bar{\alpha}^2 + \chi_{+}},
\end{equation}

since the first are of order $\mathcal{O}(\alpha^2)$ and the second are of order $\mathcal{O}(\alpha)$ (remember that $\bar{\alpha}$ is proportional to the fine structure constant $\alpha$) we have that $|r_{i}|<<|r_{sp}|$ ($i=s,p$).  Hence the integrand is positive (since the integrand has the form $I=\ln(1+A)$ where $A > 0$) and so $E_{c}(d\rightarrow 0) \rightarrow +\infty$. \\

Now we consider the limit of \textbf{large distances} ($d \rightarrow \infty$). In this limit, the reflection coefficients take the form

\begin{equation}\nonumber
r_{s} = \dfrac{1-\epsilon (0)-\bar{\alpha}^2 + \sqrt{\epsilon (0)}\chi_{-}}{1+\epsilon(0)+\bar{\alpha}^2 + \sqrt{\epsilon(0)}\chi_{+}},
\end{equation}

for the diagonal part (with a similar expression for $r_{p}$) and

\begin{equation}\nonumber
r_{sp} =  \dfrac{2\bar{\alpha}}{1+\epsilon(0)+\bar{\alpha}^2 + \sqrt{\epsilon(0)}\chi_{+}},
\end{equation}

for the off diagonal. We have defined the quantity  $\epsilon(0) \equiv 1+ \left( \frac{w_{e}}{w_{R}}\right) ^2$.\\
In this case, depending on the values of $\epsilon(0)$ different behaviours emerge. Notice, that from its definition, $\epsilon(0) \geq 1$ and so we can distinguish to extreme limits, one where $\epsilon(0)=1$ and the other with $\epsilon(0)>>1$. \\

When $\epsilon(0)=1$ one can see that we return to the previous case since the quantity $\chi_{-}$ in this limit also goes to zero. Hence $|r_{sp}|>>|r_{i}|$ ($i=s,p$) is satisfied for all distances and $E_{c}(d)$ is always positive .Therefore, using the fact that that $E_{c}(d) \rightarrow 0$ when $d \rightarrow \infty$  and that $E_{c}(d) \rightarrow +\infty$ when $d \rightarrow 0$ we deduce that there is no minimum and that the force is always repulsive. This is confirmed by the numerical calculations presented in the main text. \\


When $ \epsilon(0) >> 1$, we see that the opposite condition, $|r_{s,p}|>>|r_{sp}|$, is satisfied. Even in the worst case when $\chi_{-}$ is smallest, $\epsilon(\xi)$ in $r_{s}$ is always larger than $2\alpha$ in $r_{sp}$. The integrand at large distances is a negative quantity (since the integrand now has the form $I=\ln(1-B)$ with $B > 0$) and so $E_{c}(d)$ approaches to zero from negative values ($E_{c}(d) \rightarrow -\infty$ for $d \rightarrow \infty$).
Since at small distances $E_{c}(d)\rightarrow +\infty$ we conclude that there must be a minimum at an intermediate distance $ 0 < d_{m} < \infty$, since the function must cross the $d$ axis.\\
Notice that when $\epsilon(0)$ is strictly infinity we recover the case of an ideal metal where $r_{s,p} = \mp 1$ and $r_{sp} = 0$, with attraction at all distances.\\

To sum up, by this analytical analysis we expect the following situation to occur (when $\theta_{1}=-\theta_{2}$). As we increase $\epsilon(0)$ from one, a minimum develops at an intermediate distance $d_{m}$. This distance shifts to lower values as we increase $\epsilon(0)$ until, at $\epsilon = \infty$ we recover the ideal metal case where complete attraction occurs.\\
In the case where $\theta_{1}=\theta_{2}$ the signs inside $I$ change to give:

\begin{eqnarray}\nonumber
I = \log\left[ 1 - e^{-2 k^{(r)}_3} \left(2 r_{sp}^2+r_{p}^2+r_{s}^2\right) + e^{-4 k^{(r)}_3} \left(r_{sp}^2-r_{p}r_{s}\right)^2\right].
\end{eqnarray}

The predominant term inside the logarithm is always negative and hence the integrand is always negative. Therefore at small distances $E_{c}(d)\rightarrow -\infty$ and at large distances $E_{c}(d)\rightarrow 0$ approaching this limit from negative values, recovering attraction at all intermediate distances.\\
Finally, it can be checked by analogous methods, that the case where one $\theta$ is zero and the other one is not (dielectric - topological insulator case) results in Casimir attraction for all distances.

\subsection*{Comparison with chiral metamaterials:}

It is instructive to compare our results with the results obtainned by Zhao \emph{et al.} in \cite{Z09}. In their work, the relevant parameter to obtain repulsion is the chirality which mixes transverse electric and transverse magnetic polarizations giving in turn the off-diagonal $r_{sp}$ components. When calculating the Casimir force, by differentiation of the Lifshitz formula (Eq. (1) in the main text) they obtain the following integrand (in our notation):

\begin{equation}\label{forceEcon}
J = \dfrac{(r^2_{s}+r^2_{p}-2r^2_{sp})e^{-2kd}-(r^2_{sp}+r_{s}r_{p})^2e^{-4kd}}{1-(r^2_{s}+r^2_{p}-2r^2_{sp})e^{-2kd}+(r^2_{sp}+r_{s}r_{p})^2e^{-4kd}}.
\end{equation} 

Note that when $r_{sp}$ becomes dominant, $J$ turns negative as discussed in \cite{Z09}, enabling thus the possibility of repulsion when $r_{sp}$ is sufficiently large. In chiral metamaterials this occurs for large chirality at high frequencies (short distances) hence giving repulsion at short distances. 
The comparison with the topological insulator case is straightforward. Analytical derivation shows that the integrand for opposing signs of the topological magnetoelectric response ($\theta_{1}= -\theta_{2}$) has the exact same form as the integrand given by \eqref{forceEcon} (with properly modified reflection coefficients). In the last section we showed that, for high frequencies (short distances) $r_{sp}$ becomes dominant and hence, the topological part becomes dominant analogous to the case of large chirality in chiral metamaterials. When both topological magnetoelectric terms have equal signs ($\theta_{1}= \theta_{2}$) one obtains the same integrand with the important difference of the signs in front of the off-diagonal terms:

\begin{equation}\label{forceGru}
J = \dfrac{(r^2_{s}+r^2_{p}+2r^2_{sp})e^{-2kd}-(r^2_{sp}-r_{s}r_{p})^2e^{-4kd}}{1-(r^2_{s}+r^2_{p}+2r^2_{sp})e^{-2kd}+(r^2_{sp}-r_{s}r_{p})^2e^{-4kd}}.
\end{equation} 

As shown in the last section, even with dominant $r_{sp}$ repulsion is not possible at any frequency. Therefore the case with opposing $\theta_{1,2}$ signs is analogous to a high chirality metamaterial.\\
Note as well, that the chirality function is frequency dependent while the topological magnetoelectric response is not, since it is a topological contribution and it does not get renormalized. There is always an intrinsic high energy cut-off of the order of the lattice spacing but since we are interested in the $\mu m$ scale these effects are neglected. Hence, with opposing signs for $\theta_{1,2}$ there is always a region of repulsion for low enough distances.\\

Related to the previously discussed works there are other studies where repulsion is realised in the context of dielectric and magnetic anisotropy \cite{Rosa08a,Rosa08b,Deng08} in metamaterials (leaving chirality aside). Anisotropy can mix polarizations and thus, under the right circumstances induce repulsion as shown by \cite{Rosa08a,Rosa08b,Deng08}. The problem of finding anisotropic corrections to the Casimir force in topological insulators is not at all trivial. In the present work, we have focused on the effect of the topological magneotelectric response on the Casimir force and so we have not included the effect of anisotropy which we leave for future work. \\
For a full comparison of the cited works with the results here presented it would be necessary to include anisotropy in our study in order to draw clear conclusions. Our work is thus closer to that of chiral metamaterials, and so we leave comparison with anisotropic metamaterials for a detailed study of anisotropy. Nevertheless one could expect that since anisotropy is responsible in metamaterials for repulsion at certain distances (for some specific range of parameters), anisotropy can compete with the axionic response to develop richer behaviours. \\

\subsection*{Effect of including two oscillators:}

\begin{figure}
\includegraphics[scale=0.5]{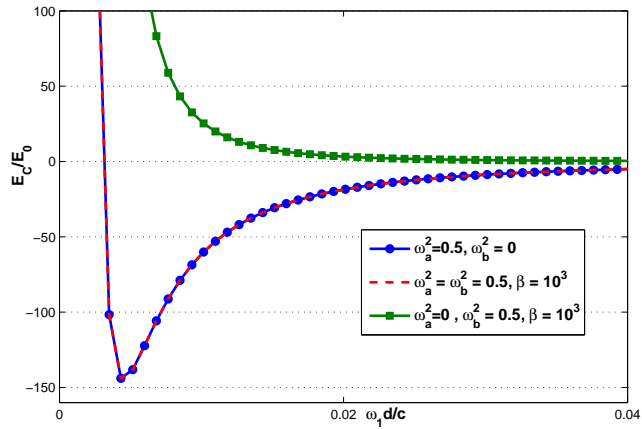}
\caption{\label{Fig:twoosc} Effect of the addition of a high frequency oscillator to the dielectric permittivity (4). The units are the same as in the figures of the main text. Green squares represent the high frequency oscillator (eq \eqref{eq:phendiel2osc} with $\omega_{a}=0$), blue circles the low frequency oscillator (eq \eqref{eq:phendiel2osc} with $\omega_{b}=0$) and the dashed red line represents the model in \eqref{eq:phendiel2osc}}
\end{figure}

In the main text we have modelled $TlBiSe_2$ with a single oscillator given experimentally by \cite{Mit92}. However in general other oscillators at different frequencies can be present when considering other topological insulators. Nevertheless, as stated above, the proof for the existence of the minimum still holds even when more oscillators are considered in (4), as long as high frequency transparency and finite zero frequency dielectric response is respected, which is in general true for insulators at low temperatures. The presence of other resonance frequencies can modify the position of the minimum just as discussed in the main text. To explore this issue a little further we have studied the case when the dielectric function is given by:

\begin{equation}\label{eq:phendiel2osc}
\epsilon(i\xi)=1+\dfrac{\omega_{a}^2}{\xi^2+ 1 +\gamma_{R}\xi}+\dfrac{\omega_{b}^2}{\xi^2+\beta^2+\gamma_{R}\xi} ,
\end{equation}

in units of $\omega_{1}$ the frequency of the first oscillator. The parameter $\beta \equiv \frac{\omega_{2}}{\omega_{1}}$ measures the position of the resonance of the second oscillator with respect to the first. To illustrate the effect of adding a high frequency oscillator we study the particular case where $\beta \sim 10^{3}$ and $\omega_{a}^2 = \omega_{b}^2= 0.5$ shown in Figure \ref{Fig:twoosc}. We have plotted the cases where both oscillators are included red dashed curve) together with the two isolated oscillators (green squares for $\omega_{a} = 0$ $\omega_{b} \neq 0$ and blue circles for $\omega_{a} \neq 0$ $\omega_{b} = 0$). Figure \ref{Fig:twoosc} shows that the effect of a high frequency oscillator is suppressed whenever $\omega_{a} \neq 0$ and it is only when $\omega_{a} = 0$ that it plays a role shifting the repulsive zone to higher distances, as expected from the analysis in the main text. Although the minimum still exists in this last case, the dielectric function is so close to one that nearly complete repulsion is obtainned, as discussed in the first section. These analysis can be summarized up as follows: 1) The minimum still exists when more oscillators are considered, in agreement with the analytical proof and 2) The lower frequency will dominate the position of the minimum since its contribution to the dielectric function will be less suppressed.

\end{widetext}

\end{document}